\documentclass[journal,transmag]{IEEEtran}
\ifCLASSINFOpdf
   \usepackage[pdftex]{graphicx}
\else
   \usepackage[dvips]{graphicx} 
\fi
\usepackage{amssymb}
\usepackage{amsmath}
\usepackage{amsfonts}
\usepackage{algorithmic}
\usepackage{algorithm}
\floatname{algorithm}{Procedure}

\usepackage{stfloats}
\usepackage{bm}
\usepackage{multirow}

\begin{document}
%
% paper title
% Titles are generally capitalized except for words such as a, an, and, as,
% at, but, by, for, in, nor, of, on, or, the, to and up, which are usually
% not capitalized unless they are the first or last word of the title.
% Linebreaks \\ can be used within to get better formatting as desired.
% Do not put math or special symbols in the title.
\title{Improvement of Resting-state EEG Analysis Process with Spectrum Weight-Voting based on LES}

% author names and affiliations
% transmag papers use the long conference author name format.

\author{\IEEEauthorblockN{
Yumeng Ye\IEEEauthorrefmark{12},
Haichun Liu\IEEEauthorrefmark{12},
TianHong Zhang\IEEEauthorrefmark{34}，
Changchun Pan\IEEEauthorrefmark{1},
Genke Yang\IEEEauthorrefmark{12},\\
JiJun Wang\IEEEauthorrefmark{34},
Robert C. Qiu\IEEEauthorrefmark{56},~\IEEEmembership{Fellow,~IEEE}}

\IEEEauthorblockA{\IEEEauthorrefmark{1}Key Laboratory of System Control and Information Processing, Ministry of Education of China,}
\IEEEauthorblockA{Shanghai Jiaotong University, Shanghai 200240, China.}
\IEEEauthorblockA{\IEEEauthorrefmark{2}Department of Automation and Shanghai Key Laboratory of Navigation and Location based Services,}
\IEEEauthorblockA{\IEEEauthorrefmark{3}Shanghai Mental Health Center, Shanghai Jiaotong University School of Medicine, }
\IEEEauthorblockA{Shanghai Key Laboratory of Psychotic Disorders, Shanghai 200030, China}
\IEEEauthorblockA{\IEEEauthorrefmark{4}Brain Science and Technology Research Center, Shanghai Jiao Tong University, Shanghai 200240, China.}
\IEEEauthorblockA{Shanghai Jiaotong University, Shanghai 200240, China.}
\IEEEauthorblockA{\IEEEauthorrefmark{5}Department of Electrical Engineering,
Shanghai Jiaotong University, Shanghai 200240, China.}
\IEEEauthorblockA{\IEEEauthorrefmark{6}Department of Electrical and Computer Engineering, Tennessee Technological University, Cookeville, TN 38505, USA.}
% <-this % stops an unwanted space
\thanks{
Corresponding author: Genke Yang (email: gkyang@sjtu.edu.cn), Changchun Pan (email: pan\_cc@sjtu.edu.cn).}}

% The paper headers
\markboth{Journal of \LaTeX\ Class Files,~Vol.1~, No.1~, January~2017}
{Shell \MakeLowercase{\textit{et al.}}: Bare Demo of IEEEtran.cls for IEEE Transactions on Magnetics Journals}
% The only time the second header will appear is for the odd numbered pages
% after the title page when using the twoside option.
%
% *** Note that you probably will NOT want to include the author's ***
% *** name in the headers of peer review papers.                   ***
% You can use \ifCLASSOPTIONpeerreview for conditional compilation here if
% you desire.

% If you want to put a publisher's ID mark on the page you can do it like
% this:
%\IEEEpubid{0000--0000/00\$00.00~\copyright~2015 IEEE}
% Remember, if you use this you must call \IEEEpubidadjcol in the second
% column for its text to clear the IEEEpubid mark.

% use for special paper notices
%\IEEEspecialpapernotice{(Invited Paper)}

% for Transactions on Magnetics papers, we must declare the abstract and
% index terms PRIOR to the title within the \IEEEtitleabstractindextext
% IEEEtran command as these need to go into the title area created by
% \maketitle.
% As a general rule, do not put math, special symbols or citations
% in the abstract or keywords.
\IEEEtitleabstractindextext{%
\begin{abstract}
Electroencephalography is a non-invasive technique for recording brain bioelectric activity, which has potential applications in various fields such as human-computer interaction and neuroscience.
Among them, analysis of the risk of schizophrenia using EEG data is a relatively new research topic.
However, there are many difficulties in analyzing EEG data, including its complex composition, low amplitude as well as low signal-to-noise ratio.
Some of the existing methods of analysis are based on feature extraction and machine learning to differentiate the phase of schizophrenia (First-episode schizophrenia, Healthy controls or Clinical high-risk) that samples belong to.
However, medical research requires the use of machine learning not only to give more accurate classification results, but also to give the results that can be applied to pathological studies.
The main purpose of this study is to obtain the weight values as the representation of influence of each frequency band on the classification of schizophrenia phases on the basis of a more effective classification method using the LES feature extraction, and then the weight values are processed and applied to improve the accuracy of machine learning classification.
We propose a method called weight-voting to obtain the weights of sub-bands’ features by using results of classification for voting to fit the actual categories of EEG data, and using weights for reclassification.
Through this method, we can first obtain the influence of each band in distinguishing three schizophrenia phases, and analyze the effect of band features on the risk of schizophrenia contributing to the study of psychopathology.
In addition, the weights applied to the original classifier can achieve the upgrade of the classification effect, which contributes to the BCI-assisted system of diagnosis and treatment.
Our results show that there is a high correlation between the change of weight of low gamma band and the difference between HC, CHR and FES. If the features revised according to weights are used for reclassification, the accuracy of result will be improved compared with the original classifier, which confirms the role of the band weight distribution.
\end{abstract}

% Note that keywords are not normally used for peerreview papers.
}

% make the title area
\maketitle

% To allow for easy dual compilation without having to reenter the
% abstract/keywords data, the \IEEEtitleabstractindextext text will
% not be used in maketitle, but will appear (i.e., to be "transported")
% here as \IEEEdisplaynontitleabstractindextext when the compsoc
% or transmag modes are not selected <OR> if conference mode is selected
% - because all conference papers position the abstract like regular
% papers do.
\IEEEdisplaynontitleabstractindextext
% \IEEEdisplaynontitleabstractindextext has no effect when using
% compsoc or transmag under a non-conference mode.

% For peer review papers, you can put extra information on the cover
% page as needed:
% \ifCLASSOPTIONpeerreview
% \begin{center} \bfseries EDICS Category: 3-BBND \end{center}
% \fi
%
% For peerreview papers, this IEEEtran command inserts a page break and
% creates the second title. It will be ignored for other modes.
\IEEEpeerreviewmaketitle

\section{Introduction}
% The very first letter is a 2 line initial drop letter followed
% by the rest of the first word in caps.
%
% form to use if the first word consists of a single letter:
% \IEEEPARstart{A}{demo} file is ....
%
% form to use if you need the single drop letter followed by
% normal text (unknown if ever used by the IEEE):
% \IEEEPARstart{A}{}demo file is ....
%
% Some journals put the first two words in caps:
% \IEEEPARstart{T}{his demo} file is ....
%
% Here we have the typical use of a "T" for an initial drop letter
% and "HIS" in caps to complete the first word.
Electroencephalography (EEG) is a method to record spontaneous bioelectric activity of the brain with a plurality of electrodes placed along the scalp.
It is a typical non-invasive technology exerting tremendous effects in the fields of brain science, human-computer interaction, and neuroscience.
EEG can be applied to personal identification because of striking intra-personal similarity and remarkable inter-personal differentiation. For example, L Ma et al proposed the individual identification technology with "brain fingerprints" based on EEG biometric\cite{Lan2015Resting}.
EEG data has a high temporal resolution.
Accordingly, EEG microstate analysis is utilized in various studies related to the Alzheimer's disease\cite{Babiloni2015Occipital}, the epilepsy\cite{Acharya2012Automated}, the REM sleep\cite{Moraes2006REM}, etc. 
One of the important applications of EEG is the diagnosis of schizophrenia\cite{Boutros2008schizophrenia}.

Schizophrenia is the clinical syndrome consisting of a group of symptoms which bring heavy medical burden to patients and society. However, the pathology of schizophrenia is still a complicated research topic. 
The current researches view schizophrenia as the syndrome originating from disruption of brain development caused by genetic or environmental factors\cite{Owen2016Schizophrenia}. 
Its diagnosis and research are based on clinical manifestations, involving the disturbance of perception, thought, facial emotion and behavior and the built-in contradiction of mental activity\cite{Joyce2007Cognitive}\cite{Barkl2014Facial}. 
In order to identify the stages of chronic schizophrenia, and to exclude the effects of long-term treatment and patient emotion, the subjects were classified as First-episode schizophrenia (FES)\cite{Bilder2000FES}, Healthy controls (HC) and Clinical high-risk for psychosis (CHR). 
CHR is defined to represent early signs and prodromal symptoms of schizophrenia\cite{Addington2011CHR}. 
A modest portion (about 29\%) of the CHR participants will convert to full psychosis after 2 years, while the other portion of them will convert to HC\cite{Fusar2012CHR}. 
Thus in the study of schizophrenia, the role of the transitional phase, CHR, cannot be ignored. However, it is not easy recognizing CHR with EEG data, resulting in obstacles to studying pathology of schizophrenia.

The differences in clinical manifestations are due to brain cell activity, which generate bioelectric activity shown in the EEG data\cite{Bera2015Noninvasive}. Thus EEG data can reflect the characteristics of various stages of schizophrenia. The traditional EEG-based research of psychosis is often focused on several distinguishing indicators such as the indicator of attention during the performance of mental tasks\cite{Harmony1996EEG} and indicators of the neurophysiological differences\cite{Duffy2015Neurophysiological}. However, schizophrenia is characterized by multiple unidentified pathogenic factors. Hence it is difficult to find certain indicators to distinguish schizophrenia phases directly, especially when CHR phase blurs the classification boundaries. 
To solve this problem, researchers propose classification methods combined with feature extraction and machine learning based on multidimensional data instead of specific indicators. M. Sabeti et al raised the approach to distinguish schizophrenic and control participants based on EEG data, which included genetic programming to select the best features and used effective classifier such as linear discriminant analysis (LDA) and adaptive boosting (Adaboost)\cite{Sabeti2011A}. 
There are many studies on the classification of two categories (FES and HC) based on EEG signals. But the study of CHR phase with resting-state EEG is a burgeoning direction. We lead in the research of classification of three categories (FES, HC and CHR) based on EEG data. 
We presented a novel classification framework consisted of linear eigenvalue statistics (LES) feature extraction (based on EEG data in the time domain or frequency domain) and machine learning to classify FES, HC and CHR. The results have shown that the accuracy of the two-category classification (FES and HC) reached 92\% and the accuracy of the three-category classification (FES, HC and CHR) reached 73\%.

In addition, the current literatures have indicated that EEG signals in the frequency domain can be divided into different frequency ranges depending on the type of brain activity. TABLE \ref{table:frequency domain} shows the commonly recognized list of significant EEG frequency bands\cite{Chan2015Systematic}.

\begin{table}[htp]
\caption{Frequency Range of major brian wave types}
\begin{center}
\begin{tabular}{cc}
\hline
Brainwave Type&Frequency Range\\
\hline
Delta&0.5 Hz to 2.75 Hz\\
Theta&3.5 Hz to 6.75 Hz\\
Low Alpha&7.5 Hz to 9.25 Hz\\
High Alpha&10 Hz to 11.75 Hz\\
Low Beta&13 Hz to 16.75 Hz\\
High Beta&18 Hz to 29.75 Hz\\
Low Gamma&31 Hz to 39.75 Hz\\
Mid-range Gamma&41 Hz to 49.75 Hz\\
\hline
\end{tabular}
\end{center}
\label{table:frequency domain}
\end{table}%

Some researchers are concerned about the role of specific EEG band. For example, EEG data of children and adolescents with autism have higher connectivity of temporal lobes with other lobes in gamma band (Sheikhani et al, 2012\cite{Sheikhani2012Detection}). The study about patients with epilepsy is focused on the alpha frequency (Larsson et al, 2012\cite{Larsson2012Alpha}). Takeuchi et al suggested the correlation of EEG low band and schizophrenia patients' third ventricular enlargement\cite{Takeuchi1994Correlation}.There are some researches representing and comparing effects of several EEG bands\cite{Bera2015Noninvasive}. Armitage analyzed the distribution of EEG bands in REM and NREM sleep stages in 1995\cite{Armitage1995The}. The analysis about EEG spectrum in the identification of phases of schizophrenia is a worthwhile research direction for studying pathology in brain and improving diagnostic quality.

In this paper, we improve the resting-state EEG analysis process using weight-voting method based on quadratic programming to find the correlation between EEG frequency bands and schizophrenia phases on the basis of the best available classification method on the strength of LES feature extraction and machine learning. It helps to study the pathology of schizophrenia through the specific internal relations between EEG frequency bands and the brain activity. It is also an innovative form of ensemble learning method to improve the accuracy of schizophrenia classification based on EEG data in comparison with existing machine learning methods.

The first section of this paper introduces the status quo of EEG and its frequency bands division, existing schizophrenia classification researches and the significance of weight allocation of EEG bands combined with machine learning method. The second section introduces the data source and the detail of the EEG band weight allocation method combined with machine learning. The third section shows the experimental results using the method mentioned in the second section. In the fourth section we summarize the method and its characteristics. The last section describes our future research directions.

\section{Data and Methods}
\subsection{Data case}
\begin{figure}[!hbt]
\centering
\includegraphics[width = 9cm]{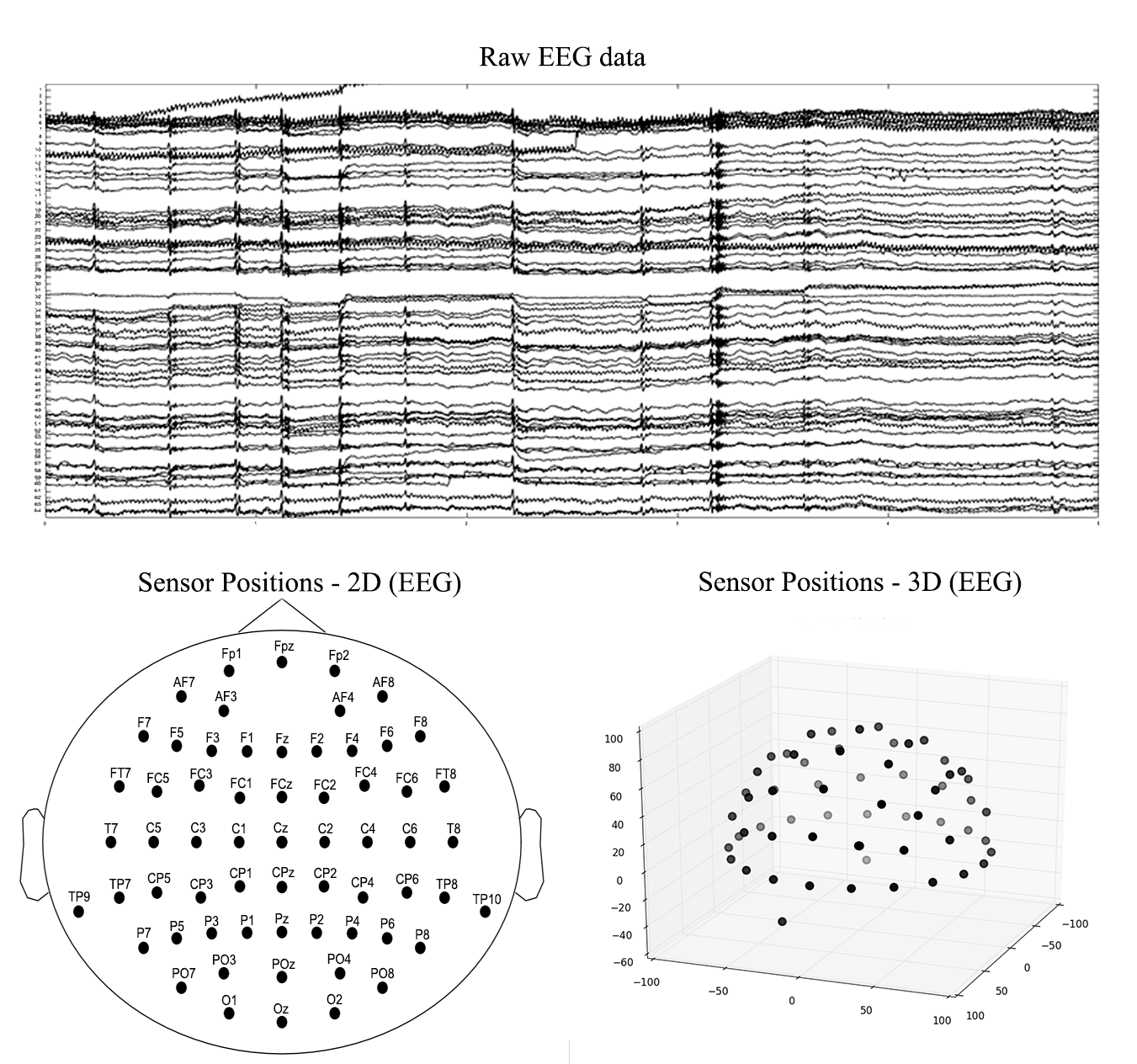}
\caption{Raw data in time domain and sensor positions of EEG channels. The 64 channels are Fp1, Fp2, F3, F4, C3, C4, P3, P4, O1, O2, F7, F8, T7, T8, P7, P8, Fz, Cz, Pz, Oz, FC1, FC2, CP1, CP2, FC5, FC6, CP5, CP6, TP9, TP10, Fpz, FCz, CPz, POz, F1, F2, C1, C2, P1, P2, AF3, AF4, FC3, FC4, CP3, CP4, PO3, PO4, AF7, AF8, F5, F6, C5, C6, P5, P6, FT7, FT8, TP7, TP8, PO7, PO8, IOLeft and IORight}
\label{fig: 64channels}
\end{figure}

The EEG data samples used for analyzing were collected from First-episode schizophrenia (FES) subjects, Healthy controls (HC) subjects and Clinical high-risk (CHR) subjects. All samples are resting-state EEG data samples taken in the same room and controlled climate with the same device. Each resting-state EEG sample consisted of 64 channels (Fig.\ref{fig: 64channels}). 

Each channel contains $t$ minutes of time-domain signals captured from specific EEG electrode. The sample frequency is 1000Hz. The actual classification labels of subjects were given by psychiatrists.

For reference convenience, TABLE \ref{table:notification} is provided to summarize some notations used throughout the paper.

\begin{table}[H]
  \centering
\caption{Notations of symbols}
\begin{tabular}{cl}
\hline
Notation&\hspace*{\stretch{1}}{Meaning}\hspace*{\stretch{1}}\\
\hline
$F$&The set of LES features. $F=\{{f_1},{f_2},\cdots,{f_m}\}$\\
$F^j$&The set of LES features of the $j$-$th$ sample.\\ &$F^j=\{{f_1^j},{f_2^j},\cdots,{f_m^j}\}$\\
$f_i$&The i-th subset of LES features.\\
$f_i^j$&The i-th subset of LES features of the j-th sample.\\
${\neg}f_i$&The complementary set of the $f_i$ in $F$. ${\neg}f_i={\complement}_{F}f_i$\\
${\neg}f_i^j$&The complementary set of the $f_i^j$ in $F^j$.\\
$m$&The number of subsets of LES features.\\
$k$&The number of samples.\\
$w_i$&The weight of ${\neg}f_i$ to the correct classification results.\\
\bm{$w$}&The vector of weight values of ${\neg}f_i$.\\
&\bm{$w$}$=({w_1},{w_2},\cdots,{w_k})^T$\\
$W_i$&The weight of $f_i$ to the correct classification results.\\
$l_i^j$&The classification label of ${\neg}f_i^j$\\
\bm{$L^j$}&The vector of classification labels of the j-th sample.\\ &\bm{$L^j}=({l_1^j},{l_2^j},\cdots,{l_m^j})$\\
$l^{j*}$&The actual label of classifications of the j-th sample.\\
$S^j$&The weighted sum of all labels for the j-th sample. \\
\bm{$L$}&The matrix contains labels of all k samples\\
&$\bm{L}=(\bm{L^1},\bm{L^2},\cdots,\bm{L^k})^T$\\
$f$&The range of EEG data concerned in frequency domain.\\
$Z$&The data blocks split from EEG data in frequency domain.\\
$B$&The number of blocks the EEG data will be divided into.\\
$\Delta f$&The sub-range of $f$ to present the range of each block. \\
&$f=\Delta f\times B$.\\
$C$&The number of channels of EEG data.\\

\hline
\end{tabular}
\label{table:notification}
\end{table}%

\subsection{Methods}
\begin{figure}[!hbt]
\centering
\includegraphics[height = 9cm]{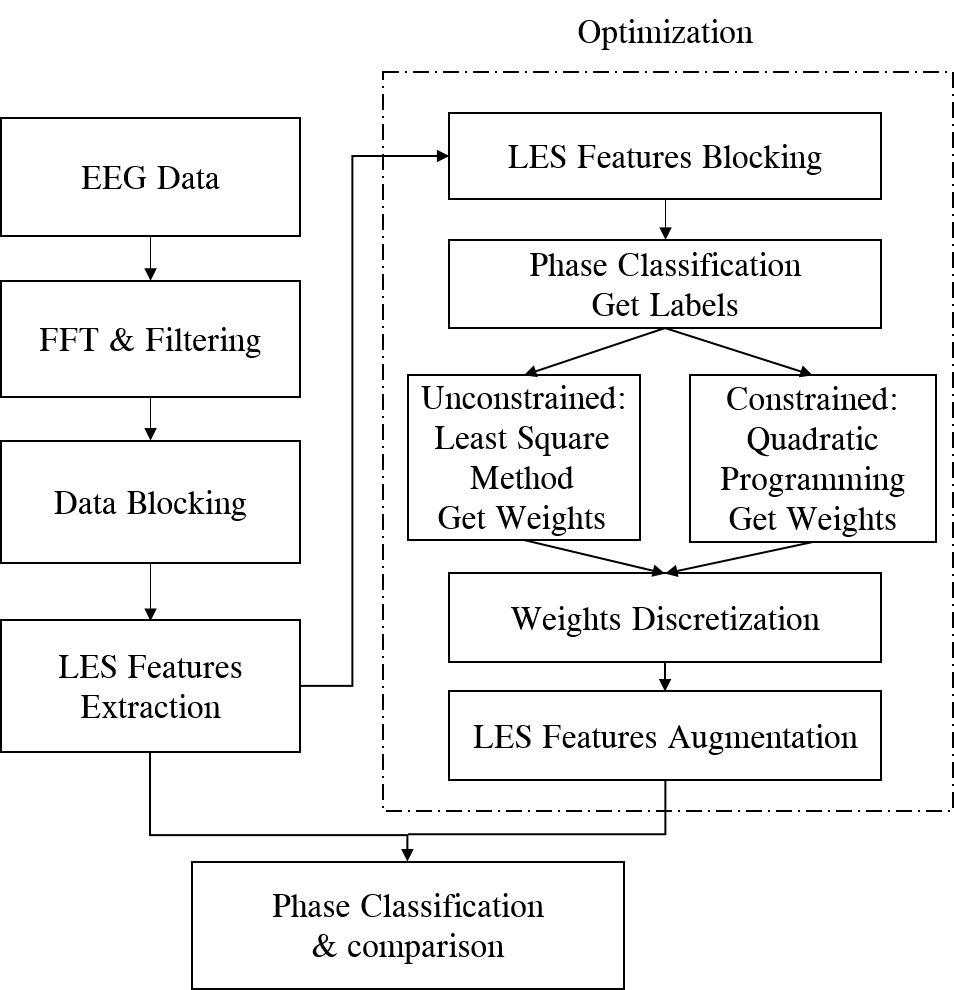}
\caption{Primary data analysis process}
\label{fig: Primary data analysis process}
\end{figure}

Primary Data Analysis Process is divided into two parts as illustrated in Fig.\ref{fig: Primary data analysis process}.

\subsubsection{Data preprocessing}
\begin{figure}[!hbt]
\centering
\includegraphics[width = 9cm]{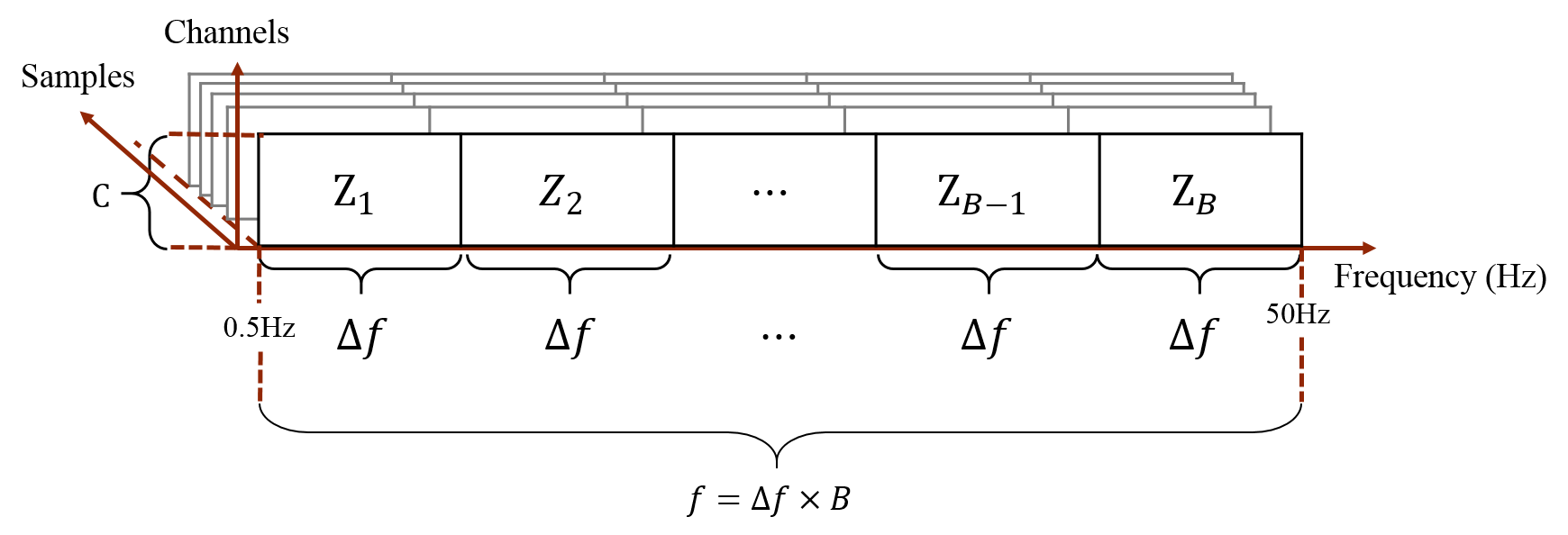}
\caption{The symbols are discribed in the TABLE \ref{table:notification}. The figure shows the preprocessing of data. The EEG data in frequency domain are split into several blocks. Each block contains the frequency signals in the range of $\Delta f$.  }
\label{fig: Preprocessing of data}
\end{figure}

Each channel of raw EEG data should be passed through a band-pass filter to remove frequencies \textless 0.5Hz and \textgreater50 Hz. This step is for removing interference factors such as noise from high-frequency power supply. The frequency bands with physiological significance in EEG data analysis appearing in TABLE \ref{table:frequency domain} are retained. Then the data would be decomposed using Fast Fourier Transform to obtain component frequencies for the further research. In order to evaluate the importance of different frequency bands, we can split the EEG spectral data into tiny data blocks shown in Fig.\ref{fig: Preprocessing of data}. Each block has the same number of channels with raw data, but its columns represent a specific sub-band of data.

\subsubsection{LES feature extraction}

In the process, the step following the data preprocessing is feature extraction. Conceptually, feature extraction is a method for transforming and extracting a set of measured values from one pattern to another to highlight representative features in distinguishing different categories and reducing dimension of massive high-dimension data. Among numerous methods, LES feature extraction combined with the subsequent classifier has the better effect on classification. It is a method to extract features from the perspective of multivariate statistics analysis based on Sample Covariance Matrix and Linear Eigenvalue Statistics (LES)\cite{Shcherbina2011Central}.

Sample Covariance Matrix deals with the question of how to approximate the actual covariance matrix on the basis of sample from the multivariate distribution. The sample covariance matrix is an unbiased and efficient estimator of the covariance matrix if the space of covariance matrix is viewed as extrinsic convex cone in $R_{p\times p}$ (Real matrix).

Consider a sample covariance matrix $M$ of the form
\begin{equation} 
M_{p\times p}=\frac{1}{n}XX^{H} 
\end{equation}

where $X$ is a matrix with p rows and n columns.

\vspace{2ex}
{\centering
\boldmath$X=$\unboldmath
$\begin{bmatrix}
 \bm{x_1}\\
 \bm{x_2}\\
 \vdots\\
\bm{x_p}\\
\end{bmatrix}$
\boldmath$=$\unboldmath
$\begin{bmatrix}
 x_{11}&x_{12}&\cdots&x_{1n}\\
 x_{21}&x_{22}&\cdots&x_{2n}\\
 \vdots&\vdots&\ddots&\vdots\\
 x_{p1}&x_{p2}&\cdots&x_{pn}\\
\end{bmatrix}$\\
}
\vspace{2ex}
 whose entries $\left \{ x_{ij} \right \}_{i=1,\ldots,p;j=1,\ldots,n}$ are independent random variables, satisfying the conditions:
$$E\left \{ x_{ij} \right \}=0,E\left \{ (x_{ij})^{2} \right \}=1$$

Let $\bm{\lambda}=\{ \lambda_{i} \}_{i=1,\ldots,p}$be eigenvalues of $M$. Then the linear eigenvalue statistic (LES), $N$, corresponding to any continuous test function $\varphi$ is
\begin{equation}
N[\varphi ]=\sum_{i=1,\ldots,p}\varphi (\lambda _{i})
\end{equation}

In big data analytics, massive random matrices can be seen as the new paradigm. And LES is considered as the analog of the Law of Large Numbers of the classical probability theory to study of the eigenvalue distribution for any ensemble of random matrices.
LES is used as the feature of data during the EEG data analysis and the classification of schizophrenia phases in previous paper\cite{Liu2017LES}. When EEG data of each sample are split into several blocks, each block could be regarded as the matrix $X$. After calculating the sample covariance matrix $M$ from $X$, the LES $N$ could be figured out. The comparison showed that the most effective test function $\varphi$ of LES used in the study was Von Neumann Entropy:
\begin{equation}
\varphi(\lambda)=-\lambda\log \lambda
\end{equation}

Thus
\begin{equation}
 N[\varphi ]=\sum_{i=1}^{p}(-\lambda_i\log\lambda_i)
 \end{equation}
Each solution of $N[\varphi ]$ serves as a feature of the sample. The number of LES features was equal to the number of blocks.

The LES features passed the significance test with the evaluation parameters far less than the traditional statistics features. In the meanwhile, LES features combined with the subsequent classifier SVM in particular resulted in the higher accuracy than other methods which tried to classify different phases of schizophrenia based on EEG data.

Because of the superiority of LES features, we adopt LES feature extraction in following steps of the process shown in Fig.\ref{fig: Primary data analysis process} where the features are used for representing the bands and for further classification.

\subsubsection{Quadratic Programming Weight-Voting Combination}

After extracting n features with LES feature extraction to constitute the set $F$, we consider evaluating the contribution (i.e. weight) of different EEG bands on the accuracy of classification for two purposes. (1)The contribution of different EEG bands on distinguishing phases of schizophrenia could be reflected in the physiological activities of the human brain, which is the important research approach of the pathogenesis of schizophrenia\cite{Bera2015Noninvasive}. (2)The revision of LES features based on the contribution of EEG bands leads to the improvement of classification effect. This is also the verification of the weights. For these reasons, we propose a process mode derived from weighted-voting method of ensemble learning\cite{Dietterich2000Ensemble}. 

In this process mode, we depart the features into several regions and set weak classifiers with sub-features. These weak classifiers are combinated using weighted-voting and quadratic programming fitting for calculating weight values of features of EEG bands. To verify the effect of the weight values, we should remain the machine learning method unchanged. Thus we use the weight values to revise the original LES features instead of changing the structure of classification algorithm such as using these values in the ensemble classifier directly. 

More specifically, the set $F$ is equally divided into m subset $f_i$, which represents the features of the specific EEG band.
Each subset $f_i$ contains $n/m$ LES features. Due to the small number of features in $f_i$, the average accuracy of classification using machine learning based on $f_i$ is hovering around 50\%. It means that the classification results are almost random and not reliable.
In order to construct a reliable classifier to obtain the weight of different EEG bands by quadratic programming, we build a structure similar to the stepwise regression and use the backward elimination in the original features.\cite{Franke2010Stepwise}. In the structure, we consider using ${\neg}f_i$, the complementary set of $f_i$ in $F$, as a corresponding sub-feature-set to construct each weak classifier. 
Then we get the classification label set consisting of all the classification labels $l_i$ of ${\neg}f_i$ with these weak classifiers.The result of the deletion of each $f_i$ shows its statistically insignificant deterioration of the classifier. If the $f_i$ is insignificant in classification, the label $l_i$ should be the same as the actrual label of the sample in most tests.

According to the weighted-voting method, we assign the weight called $w_i$ to the label of ${\neg}f_i$, then obtain the weighted sum of all labels called $S^j$.
\begin{equation}
w_1l_1^j + w_2l_2^j+\cdots+w_ml_m^j= S^j
\end{equation}
The labels and weights can be seen as vectors.\\

{\centering
$\bm{w=}$
$
\begin{bmatrix}
 w_1\\
 w_2\\
 \vdots\\
 w_m\\
\end{bmatrix}
$
,
$\bm{S=}$
$
\begin{bmatrix}
S^1\\
S^2\\
\vdots\\
S^k\\
\end{bmatrix}$,
\boldmath$l^*=$\unboldmath
$\begin{bmatrix}
l^{1*}\\
l^{2*}\\
\vdots\\
l^{k*}\\
\end{bmatrix}$,\\
}
\quad
\\{ }

{\centering
\boldmath$L=$\unboldmath
$\begin{bmatrix}
 \bm{L^1}\\
 \bm{L^2}\\
 \vdots\\
\bm{L^k}\\
\end{bmatrix}$
\boldmath$=$\unboldmath
$\begin{bmatrix}
 l_1^1&l_2^1&\cdots&l_m^1\\
 l_1^2&l_2^2&\cdots&l_m^2\\
 \vdots&\vdots&\ddots&\vdots\\
 l_1^k&l_2^k&\cdots&l_m^k\\
\end{bmatrix}$\\
}

\vspace{2ex}
Thus according to the equation(3), we have
\boldmath
\begin{equation}
 Lw = S
\end{equation}
\unboldmath

In the ideal condition, the weighted sum $S^j$ should be equal to the actual label of the sample $L^{j*}$, i.e. $S^j = L^{j*}$.The equations of each sample form an overdetermined equation system, which can be solved with the least square method\cite{Lawson1974Solving}.

It means that we need to minimize the sum of squared residuals of equations. The objective function of the problem is

\begin{equation}
  \begin{aligned}
  minf(\bm{w})&=\frac{1}{2}\left\|{\bm{Lw-l^*}}\right\|^2\\
   &=\frac{1}{2}\sum_{j=1}^k{\bm{(L^jw-l^{j*}})^2}
   \end{aligned}
\end{equation}
Because it is the unconstrained optimization problem, the minimum of objective function can be reached by making the gradient equal to zero, as shown in equation (8).
\begin{equation}
  \frac{\partial f(\bm{ w })}{\partial\bm{w}}=\bm{L^TLw-L^Tl^*}=0
\end{equation}
As \bm{$L^TL$} is a nonsingular matrix, the solution $\bm w$ of the problem from equation (8) is
\begin{equation}
  \bm{w=(L^TL)^{-1}L^Tl^*}
\end{equation}

When the least square method is used in the fitting process that contains noise, there may be over-fitting phenomena because the learning model is too complicated for the training sample. In order to control the complexity of the model, we consider the least square fitting with constraints. The way we add constraints to the least square model is based on the classification results derived from the classifier. First, we exclude the negative weight caused by misjudgment, thereafter we normalize the weight. For this reason, we add the equation constraint and the inequality constraint to the original problem to transform the problems into quadratic programming problems with constant terms as shown in equation (10).

\begin{equation}
  \begin{aligned}
  min&\   f(\bm{w})=\frac{1}{2}\left\|{\bm{Lw-l^*}}\right\|^2\\
  s.t.&\quad\ \ \bm{w}\geqslant 0\\
  &\sum_{i=1}^mw_i=1
   \end{aligned}
\end{equation}

 Since this quadratic programming problem is a small-scale problem, we use the most efficient algorithm for this kind of problems called Active Set Algorithm\cite{Nocedal2006Numerical} to solve it. The Active Set Algorithm for quadric programming is improved from the Simplex Algorithm for linear programming. 
 This type of algorithm is characterized by the iterative process in which the objective point should move along the boundary of the constraints until it reaches the optimal point of the problem. The Active Set means the set containing the inequality constraints(with $\geq$ and $\leq$ in the constraints) whose equality holds at the current solution point during the quadratic programming. The Active Set Algorithm begins by assuming an initial active set which is called Working Set, and the optimal solution of the quadratic programming subproblem is solved on the basis of these equality constraints. In the process of solving the optimal solution of the subproblem, the constraint is added to the working set according to the step-size parameters. When the optimal solution of the subproblem is obtained, we should judge whether it is the optimal solution of original quadratic programming. If there are some constraints with negative Lagrange multipliers in the working set, we need to remove these constraints from the working set and solve new subproblem. The process of the algorithm will repeat until all the constraints in the working set have nonnegative Lagrange multipliers and the corresponding solution is the optimal solution of the original quadratic programming problem.

 Using the Active Set Algorithm to solve the quadratic programming problem, we calculate the weight value $w_i$, which refers to the contribution of each ${\neg}f_i$ to the correct classification result. In order to get the actual influence of different EEG bands instead of its complementary sets, we devide each $w_i$ to m-1 feature segments corresponding to ${\neg}f_i$. Because each $w_i$ presents the contribution of m-1 feature segments in ${\neg}f_i$, the $w_i$ is equally allocated to these segments. Thus the weight of each $f_i$ is superimposed m-1 assigned values of $w_i$ from other feature segments. For example, if there were m subsets in $F$ and we got $w_1$ to $w_m$ with Least square method, the actual weight of $f_1$ would be $$W_1=(w_2+w_3+\cdots+w_m)/(m-1)$$ and the weight of $f_2$ would be $$W_2=(w_1+w_3+w_4+\cdots+w_m)/(m-1),$$ and so on for each weight of the subsets.

\subsubsection{Postprocessing for classification}

According to the upper and lower limits and the resolution of the weight values, we discretize the weights into multiple levels. After calculating the weight levels in the previous step, we use them to construct new feature set by means of feature augmenting\cite{Xu2015A}. That is, the original features in one subset who share the same weight are replicated for several times and the number of time is in proportion to the weight levels. Finally, these augmented features can be used for machine learning classification which is also used in weak classifiers. 

The importance of different frequency bands in the classification of schizophrenia can be analyzed by this method. At the same time, since the result of the quadratic programming fitting is closest to the exact classification, the revised feature set based on the weights leads to an improvement in accuracy. We designed the experiment to verify the practical effect of this method in classification.

 \begin{figure*}[!hbt]
 \centering
 \includegraphics[width=18cm]{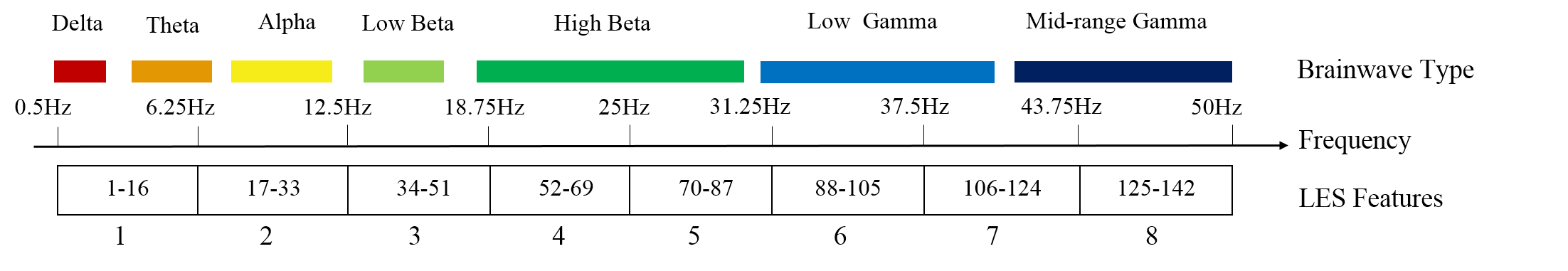}
 \caption{Comparision among bands, features and frequency}
 \label{fig: bandvalue}
 \end{figure*}

\section{Experiment}

The experiment is divided into two cases, which are the classification test for two-category (FES, HC) and the classification test for three-category (FES, HC, and CHR).  Each case has three steps: (1)preprocessing and LES feature extraction, (2)weight calculation, and (3)classification with augmented features based on weight. 

The actual EEG data we used in the experiments contained 120 subjects (40 subjects of HC, 40 subjects of FES and 40 subjects of CHR). In the first step, there are 10 samples in each of three categories being selected randomly as the test samples and the other 30 samples in the same category as training samples. That is, in the three categories classification training samples, the numbers of FES, HC and CHR samples are 30, 30 and 30. In the case of two categories classification training samples, the numbers of FES and HC samples are 30 and 30. The test of each case will be repeated 20 times containing random selection and cross validation to ensure the credibility of the experiment. After the preprocessing process containing filtering, fast Fourier transform, and calculation of power spectral density of each channel, EEG data becomes the PSD matrix with 64 rows and 14200 columns. Then in the process of feature extraction by LES method, the PSD matrix is divided into 142 matrices of 64 * 100 by columns. These matrices are arranged from low to high according to the corresponding frequency band. The linear eigenvalue statistics of each matrix are calculated based on the method proposed in the previous works which using Von Neumann Entropy as the test function on the eigenvalues of sample covariance matrix of EEG data blocks. The feature extraction result is a vector of 142 LES values corresponding to each band shown in Fig.\ref{fig: bandvalue}.

In the second step, 142 LES features are divided into eight parts, and the weights are calculated according to the constrained quadratic programming and non-constrained least square method.

Finally, the original LES features and the revised features obtained by the weight discretization and the feature augmentation method described in the previous section are tested through two machine learning algorithms, SVM and KNN. Because the kernel of SVM and KNN are not weight-adjusted and the effect of the two algorithms are better than other algorithms before feature revision. The results of experiment are shown in the next section.

\section{Results and discussion}
\subsection{The weight of frequency band}
\begin{figure}[!hbt]
\centering
\includegraphics[width = 9cm]{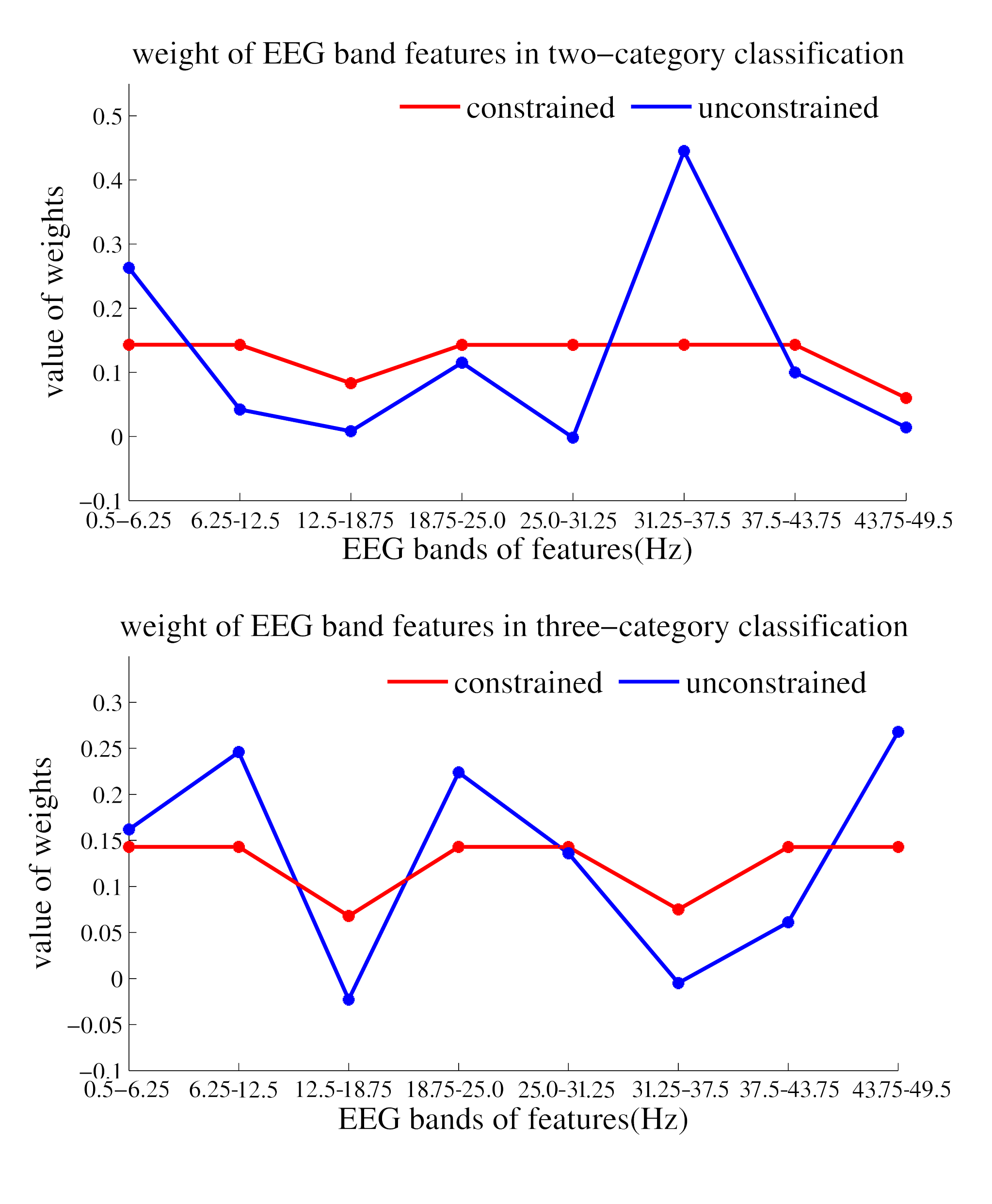}
\caption{weight values of features of frequency bands under different methods}
\label{fig: weights}
\end{figure}

The result under constrained quadratic programming shows that the frequency bands with the least impact on the classification of two-category are 12.5Hz-18.75Hz and 43.75Hz-50Hz, that is, the low beta and most of the Mid-range Gamma have no significant effect on the classification of FES and HC. There is similar test result from the non-constrained condition. Moreover it presents the significant role of 0.5Hz-6.25Hz and 31.25Hz-37.5Hz bands which almost cover Delta, Theta and Low Gamma bands.

For the three-category classification, the weights obtained with constrained condition shows that 12.5Hz-18.75Hz and 31.25Hz-37.5Hz, i.e. Low Beta and Low Gamma bands, have less influence on the correct classification result. The results of the unconstrained method also confirm that the effects of Low Beta and Low Gamma bands are very weak compared to Alpha, High Beta and Mid-range Gamma bands’ effects in the three-category classification.

It is shown by the results that the more obvious difference between the two-category and the three-category classification are mainly reflected in the 31.25Hz-37.5Hz band (most part of Low Gamma band). The three-category classification test includes CHR samples, which blurs the boundaries between the HC and the FES samples. Thus the distinction between the weights of the two-category and the three-category classification means that CHR has a strong effect of confounding in the 31.25Hz-37.5Hz band. This indicates that in transition phase from HC to FES, low gamma band has more intense and continuous transformation. Furthermore, brainwave characteristics of CHR samples are mainly concentrated in Alpha, High Beta and Mid-range Gamma who have a greater influence in the classification.

\subsection{The results of classification}
In order to verify the results of the weight calculation, and to improve the accuracy of classification, we do discretization for weight values, and revise the features by augmentation.

The definition of accuracy is the ratio between the number of correct labels given by classifier and the number of labels of all samples:
$$ Accuracy=\frac{correct\ labels}{all \ labels}\times100\% $$
And in two-category classification test, the sensitivity (true positive ratio) and specificity (true negative ratio) are calculated with the following equations:
$$ Sensitivity=\frac{TP}{TP+FN}\times100\% $$
$$ Specificity=\frac{TN}{TN+FP}\times100\% $$
where TP = true positive; TN = true negative; FP = false positive; FN = false negative. (We set HC as Positive label and FES as Negative label.)

\begin{figure}[!hbt]
\centering
\includegraphics[width = 9cm]{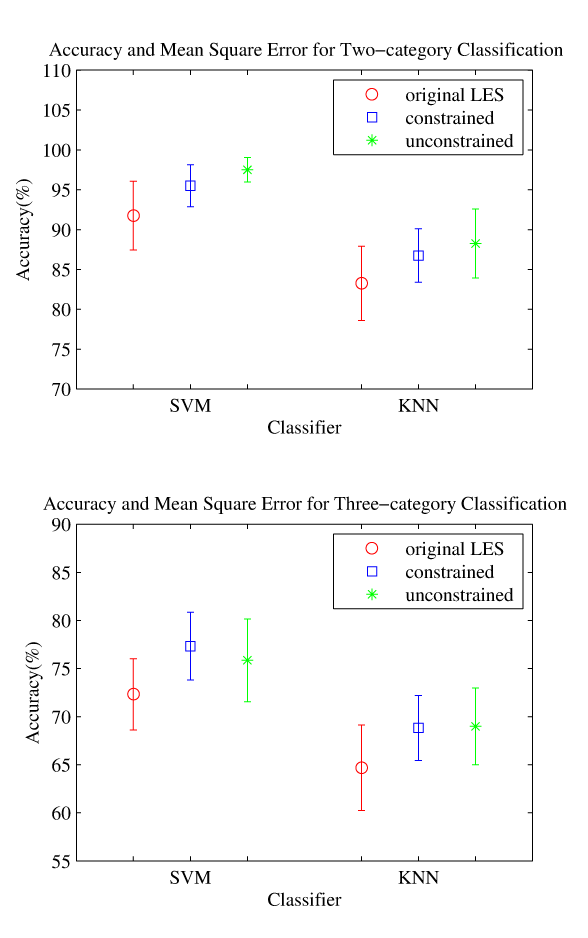}
\caption{The mean accuracy and mean square error of two-category and three-category classification under different classifiers and algorithms. Original LES: LES features without optimization, constrained: under algorithm with constrained quadratic programming, unconstrained: under algorithm with unconstrained least square method.  }
\label{fig:Accuracy}
\end{figure}

The final statistical results are reflected in the average and mean square error of accuracy of 20 cross validation tests shown in TABLE \ref{table: effect of different classifiers} and Fig.\ref{fig:Accuracy}.

\begin{table}[htp]
\caption{Accuracy and mean square error of different classification conditions}
\begin{center}
\begin{tabular}{clcc}
\hline
Classifier&Algorithm&HC\&FES&HC\&FES\&CHR\\
\hline
&original LES&$91.75\pm 4.32\%$&$72.33\pm 3.69\%$\\
SVM&constrained&$95.50\pm 2.62\%$&$77.33\pm 3.52\%$\\
&unconstrained&$97.50\pm 1.54\%$&$75.83\pm 4.29\%$\\
\hline
&original LES&$83.25\pm 4.67\%$&$64.67\pm 4.45\%$\\
KNN&constrained&$86.75\pm 3.35\%$&$68.83\pm 3.38\%$\\
&unconstrained&$88.25\pm 4.32\%$&$69.00\pm 3.99\%$\\
\hline
\end{tabular}
\end{center}
\label{table: effect of different classifiers}
\end{table}%

The results shown in Fig.\ref{fig: weights} demonstrate that the classification accuracy of the treatment has a certain improvement.

When we use the SVM classifier, the average accuracy of two-category classification with the features improved by constrained quadratic programming and unconstrained least square reach 95.5\% and 97.5\% respectively, which is increased by 3.75\% and 5.75\% compared with the accuracy of the same test with the original LES feature. The accuracy of the same test with the KNN algorithm is 83.25\% from the untreated LES feature, 86.75\% from the features improved by constrained quadratic programming and 88.25\% from the features improved by unconstrained least square. It can be seen that the classification accuracy has been greatly improved. The weights obtained by the unconstrained least square leads to a more pronounced effect on the classification comparatively. For the two-category classification, the features of the Low Gamma band also contributes to distinguish between HC and FES. The high value of weights calculated by unconstrained means an increasing error probability of the classification after removing the corresponding band. Thus the single frequency band could heighten the accuracy of classification.

\begin{table}[htp]
\caption{Two-category classification results by different methods}
\begin{center}
\begin{tabular}{cccc}
\hline
Classifier&Accuracy&Sensitivity&Specificity\\
\hline
LDA&$85.90\%$&$86.45\%$&$85.29\%$\\
Adaboost&$91.94\%$&$92.91\%$&$90.57\%$\\
\hline
SVM&$97.50\%$&$98.50\%$&$96.50\%$\\
KNN&$88.25\%$&$89.00\%$&$87.50\%$\\
\hline
\end{tabular}
\end{center}
\label{table: effect of different classifiers from Sabeti}
\end{table}%

The results are also compared with the results from the approach proposed by M. Sabeti, which presented the effect of geniric programming feature extraction and the classifiers using LDA and Adaboost. The results (TABLE \ref{table: effect of different classifiers from Sabeti}) in Sabeti's research showed that the accuracy of two-category classification with LDA method was 85.9\% and the accuracy with Adaboost was 91.94\%\cite{Sabeti2011A}. The average accuracy in our work using the SVM classifier can reach 97.5\% in unconstrainted condition, having a 5.56\% higher accuracy. This illustrates the superiority of the process containing LES feature extraction, SVM classifier and the optimization methods based on serching for significant features of EEG bands.
In two-category classification, the sensitivity values of SVM and KNN are slightly higher than whose specificiy values. It means HC has more distinguishing features in classification. 

For three-category classification, the accuracy shown in TABLE \ref{table: effect of different classifiers} with SVM algorithm from original LES features is 72.33\%. The average accuracy of classification using unconstrained least square reaches 75.83\%.  The number reaches 77.33\% using constrained quadratic programming. The revised features bring a 5.00\% improvement in accuracy. Compared to the accuracy (64.67\%) from original features with KNN algorithm, the result with optimization method is increased by 4.16\% and 4.33\% respectively. The optimization method of the weight allocation has also significantly improved the accuracy of the three-category classification. While there is no further improvement of result by unconstrained method compared to that by constrained method as in the case of two categories. The reason is shown in Fig.\ref{fig: weights}. Since the weights obtained from the unconstrained least square are fluctuating observably, the bands other than the Low Beta and Low Gamma bands both have relatively high value with two method. When the weights are applied to augment of features, they have similar trends and lead to similar proportion of numbers of features based on the principle of augment. It results in the similar average accuracy of classification.

From the results, the mean square error of the experimental results fluctuates slightly around 3.5\%. The smaller mean square error indicates that the experimental results are more concentrated, which presents the stablity of experimental results.

In order to further analyze the classification results among three categories, the statistics for the accuracy of each category are necessary steps. The TABLE \ref{table: accuracy of classification} shows the three-category classification results of each category in constrained conditions. Test label means the result given by classifiers, actual label means the standard of the classification. Each value in the table refers to the average ratio of the event that test result matches the actual label. The values reflect a strong ability of two classifiers to identify FES. The percentages of correct classification reach 91.50\% and 87.00\%, which match the values in two-category classification. While the values of HC and CHR show some degree of confusion of classifiers to distinguish HC and CHR. There are 27.00\% of CHR participants being regarded as HC and 25.50\% of HC participants being mistaken for CHR, which indicate considerable similarity between the features of HC and CHR participants. The test for KNN classifier gets an analogous result. It confirms the tendency for CHR phase to convert into HC phase in future\cite{Fusar2012CHR}.

\begin{table}[htp]
\caption{Three-category classification results of each category}
\begin{center}
\begin{tabular}{c|c|ccc}
\hline
&&&&\\
Classifier&Test label&&Actual label&\\
\cline{3-5}
&&HC&FES&CHR\\
\hline
&HC&$72.00\%$&$6.00\%$&$27.00\%$\\
SVM&FES&$2.50\%$&$91.50\%$&$4.00\%$\\
&CHR&$25.50\%$&$2.50\%$&$69.00\%$\\
\hline
&HC&$63.00\%$&$6.00\%$&$33.00\%$\\
KNN&FES&$6.50\%$&$87.00\%$&$7.50\%$\\
&CHR&$30.50\%$&$7.00\%$&$59.50\%$\\
\hline
\end{tabular}
\end{center}
\label{table: accuracy of classification}
\end{table}%

\section{Conclusion}
In this study, we proposed a method with three steps to find a group of weight values representing the importance of each frequency band’s features based LES. First step is dividing the EEG data into blocks according to the frequency band and classifying the various phases of schizophrenia based blocks respectively with the reliable classifier. Second step is establishing a voting combination of multiple classification results with weight values to fit the actual labels. Final step is using Least Square Method and Active Set Algorithm to obtain the best fitting weights.

By analyzing the weights, it can be concluded that the differences in EEG energy features between HC and FES are concentrated in the delta, theta and low gamma bands, while the low beta and most of the mid-range gamma bands do not show a significant difference. When focusing on CHR, the difference in weights between three-category and two-category classification is mainly reflected in the low gamma band. The features of this band brings confusions between CHR-HC or CHR-FES. This finding can be used as a reference for studying the transitional phase of schizophrenia through the low gamma band-related brain cell activity, which helps to study the pathogenesis of schizophrenia.

On this basis, we study the impact of weight on the classification of schizophrenia phases. When the values of weights is used to augment the original features in accordance with the proportion and the same machine learning methods are still used to classify different phases, the classification results are improved compared with original features. Through the cross validation, the average classification accuracy is raised by 4\% to 5\%. Moreover, we find that in the two-category classification, the unconstrained least square method achieves higher accuracy than the constraint method. When the individual band weight values are highlighted, the unconstrained method results in an enlargement of the effect in the feature selection, which leads to an enhancement in accuracy. It means that the low gamma band has an important effect on the classification.

The weight-voting method is constituted with the process that contains extracting LES frequency domain features, building classifier to vote for weights and reconstructing features to classification. Because of the extensive application of EEG data, study of human brain activity from the frequency domain is a developing research direction. This method is extensible, which can be extended to apply in other related fields such as the analysis of brain cell activity and brain diseases based on frequency domain EEG data.

\section{Future directions}
There is extensive research space for EEG data analysis of human psychiatric disorders. Our current work is focused on EEG spectral analysis. The method for frequency band division adopted in our study is largely consistent with the existing brainwave division way, and some specific bands with significant contribution to classification of schizophrenia phases are obtained. 
The prospective work on EEG data analysis could be two major directions including the further data analysis based on frequency domain and the data analysis based on spatial information of EEG. 
From the perspective of frequency domain, we have found the inequality of bandwidth of brainwaves. In order to improve the accuracy of results from calculating importance of frequency bands and from classification of schizophrenia phases, we can subdivide the frequency bands. After the division, the weight value of a segment will refer to more targeted frequency band. It helps to draw a detailed contribution map of frequency bands.
From the perspective of spatial information, we believe that significant advantages of EEG lie in its sampling method compared with other BCI detection methods. The channels of EEG data represent the spatial position and distribution of brainwaves. Thus the EEG data could be used to obtain brain-structure-related conclusions. Therefore, we will combine the relevance of EEG channels with machine learning and assess the importance of the channels for the diagnosis of psychiatric disorders and the study of pathology based on data analysis.

\section{ACKNOWLEDGMENT}
We are appreciated for department of EEG source imaging leaded by professor Jijun Wang and professor Chunbo Li from SHJC for data providing and discussion. We also grateful to Dr. Tianhong Zhang from SHJC for his expert collaboration on data analysis.

This work is supported by National Natural Science Foundation(NNSF) of China under Grant 61290323 and Nature Science Foundation of Shanghai under 16ZR1416500.

% if have a single appendix:
%\appendix[Proof of the Zonklar Equations]
% or
%\appendix  % for no appendix heading
% do not use \section anymore after \appendix, only \section*
% is possibly needed

% use appendices with more than one appendix
% then use \section to start each appendix
% you must declare a \section before using any
% \subsection or using \label (\appendices by itself
% starts a section numbered zero.)
%

%\appendices
%\section{Proof of the First Zonklar Equation}
%Appendix one text goes here.

% you can choose not to have a title for an appendix
% if you want by leaving the argument blank
%\section{}
%Appendix two text goes here.

% use section* for acknowledgment
% \section*{Acknowledgment}

% The authors would like to thank...

% Can use something like this to put references on a page
% by themselves when using endfloat and the captionsoff option.
\ifCLASSOPTIONcaptionsoff
  \newpage
\fi

% trigger a \newpage just before the given reference
% number - used to balance the columns on the last page
% adjust value as needed - may need to be readjusted if
% the document is modified later
%\IEEEtriggeratref{8}
% The "triggered" command can be changed if desired:
%\IEEEtriggercmd{\enlargethispage{-5in}}

% references section

% can use a bibliography generated by BibTeX as a .bbl file
% BibTeX documentation can be easily obtained at:
% http://mirror.ctan.org/biblio/bibtex/contrib/doc/
% The IEEEtran BibTeX style support page is at:
% http://www.michaelshell.org/tex/ieeetran/bibtex/
%\bibliographystyle{IEEEtran}
% argument is your BibTeX string definitions and bibliography database(s)
%\bibliography{IEEEabrv,../bib/paper}
%
% <OR> manually copy in the resultant .bbl file
% set second argument of \begin to the number of references
% (used to reserve space for the reference number labels box)

% \begin{thebibliography}{1}

% \bibitem{IEEEhowto:kopka}
% H.~Kopka and P.~W. Daly, \emph{A Guide to \LaTeX}, 3rd~ed.\hskip 1em plus
%   0.5em minus 0.4em\relax Harlow, England: Addison-Wesley, 1999.

% \end{thebibliography}

\bibliographystyle{IEEEtran}
\bibliography{eeg}
% biography section
%
% If you have an EPS/PDF photo (graphicx package needed) extra braces are
% needed around the contents of the optional argument to biography to prevent
% the LaTeX parser from getting confused when it sees the complicated
% \includegraphics command within an optional argument. (You could create
% your own custom macro containing the \includegraphics command to make things
% simpler here.)
%\begin{IEEEbiography}[{\includegraphics[width=1in,height=1.25in,clip,keepaspectratio]{mshell}}]{Michael Shell}
% or if you just want to reserve a space for a photo:

% \begin{IEEEbiography}{Yumeng Ye}
% Biography text here.
% \end{IEEEbiography}

% \begin{IEEEbiography}{Haichun Liu}
% Biography text here.
% \end{IEEEbiography}

% \begin{IEEEbiography}{Changchun Pan}
% Biography text here.
% \end{IEEEbiography}

% \begin{IEEEbiography}{Genke Yang}
% Biography text here.
% \end{IEEEbiography}

% \begin{IEEEbiography}{Robert C. Qiu}
% Biography text here.
% \end{IEEEbiography}

% if you will not have a photo at all:
%\begin{IEEEbiographynophoto}{Robert C. Qiu}
%Biography text here.
%\end{IEEEbiographynophoto}

% insert where needed to balance the two columns on the last page with
% biographies
%\newpage

% You can push biographies down or up by placing
% a \vfill before or after them. The appropriate
% use of \vfill depends on what kind of text is
% on the last page and whether or not the columns
% are being equalized.

%\vfill

% Can be used to pull up biographies so that the bottom of the last one
% is flush with the other column.
%\enlargethispage{-5in}

% that's all folks
\end{document}